# HEAT LOADS AND CRYOGENICS FOR HE-LHC

D. Delikaris, L. Tavian, CERN, Geneva, Switzerland

*Abstract*

We report preliminary considerations on cryogenics for a higher-energy LHC ("HE-LHC") with about 16.5 TeV beam energy and 20-T dipole magnets. In particular we sketch the heat loads scaled on the proposed principal beam parameters and size the cryogenic plants for different operating temperature of the beam screens.

## INTRODUCTION

Similar to the LHC, the heat deposited in the HE-LHC will reach 3 different temperature levels:

- the thermal shield temperature level (TS) between 50 and 75 K,
- the 5-K heat intercept (HI) and beam screen temperature level (BS) between 4.6 and 20 K (40-60 K or 85-100 K as an alternative compatible with vacuum specification), and
- the cold mass temperature level (CM) at 2 K.

It is also assumed that specific cryogenic systems will be needed for insertion magnets and RF cavities. These insertions are not defined yet; consequently, in the following, only the continuous cryostats (CC) will be considered, i.e. arcs plus dispersion suppressors and associated current feed boxes.

## HEAT INLEAKS

In first approximation the thermal performance of the HE-LHC cryostat is assumed to be similar to the one of the LHC cryomagnet. In addition, it is assumed that the LHC cryoline (QRL) is used with its present thermal performance [1]. The specific heat inleaks are given in Table 1.

Table 1: Specific heat inleaks on magnets and cryoline

| Temperature level | | LHC | HE-LHC |
|---|---|---|---|
| TS (50-75 K) | [W/m] | 7.7 | 7.7 |
| HI (4.6 K) | [W/m] | 0.23 | 0.23 |
| CM (2 K) | [W/m] | 0.21 | 0.21 |

## RESISTIVE HEATING IN SUPERCONDUCTING SPLICES

For HE-LHC, a main magnet current of 18 kA is assumed. The resistive heating in the magnet splices is proportional to the square of the magnet current, to the splice electrical resistance and to the number of splices. The corresponding heat load is deposited at the CM temperature level. As the HE-LHC hybrid coil design is based on 3 different cables, it is assumed that the number of splices increases by a factor 1.5 with respect to the LHC coil design based on 2 different cables. Table 2 gives the main parameters related to the resistive heating. The increase of the magnet current and of the number of splices, both by 50 %, translates into an increase of the resistive heating by a factor 3.4 with respect to the nominal LHC.

Table 2: Resistive heating in magnet splices

| | LHC nominal | HE-LHC |
|---|---|---|
| Main magnet current [kA] | 12 | 18 |
| Splice resistance [nΩ] | 0.5 | 0.5 |
| Number of splice per arc [-] | 2500 | 3750 |
| Resistive heating on CM [W/m] | 0.1 | 0.34 |

## CURRENT LEAD COOLING

Concerning the cooling of the current leads (CL), it is assumed that HE-LHC is using the same type of HTS current lead as the LHC with the same cooling performance, i.e. a specific cooling rate per kA of 54 mg/s of helium between 20 and 300 K. In addition, as the optics of the HE-LHC is not yet fully defined the number of individually powered magnets is not known; consequently, it is assumed that the total current entering or exiting is proportional to the main magnet current. In addition, as for the LHC, it is assumed that high-load sectors enter two times more current than low-load sectors. Table 3 lists the main parameters for the current lead cooling.

Table 3: Current lead cooling

| | LHC nom. | HE-LHC |
|---|---|---|
| Main magnet current [kA] | 12 | 18 |
| Total current in/out [kA] | 2750 | 4130 |
| Total current high load sector CC [kA] | 460 | 690 |
| Total current low load sector CC [kA] | 230 | 345 |
| Specific CL cooling flow [mg s$^{-1}$ kA$^{-1}$] | 54 | 54 |
| High-load sector CL cooling flow [g/s] | 25 | 37 |
| Low-load sector CL cooling flow [g/s] | 12 | 19 |

## BEAM-INDUCED LOADS

The parameters impacting the beam-induced loads are the beam energy, the bunch population, the number of bunches, the bunch length and the beam-screen aperture. Table 4 gives the scaling laws to be applied for the different beam-induced loads. Table 5 lists the parameters and the beam-induced loads for the nominal LHC and the

HE-LHC. Compared with the nominal LHC, all beam-induced loads on the beam screens increase for the HE-LHC. The biggest change concerns the synchrotron-radiation load, which increases by a factor 17.

Table 4: Scaling laws of beam induced heat loads

| Beam-induced load | Energy E | Bunch population $N_{bunch}$ | Bunch number $n_{bunch}$ | Bunch length $\sigma_z$ [rms] | Beam-screen aperture b | Temp. level |
|---|---|---|---|---|---|---|
| Synchrotron radiation | $E^4$ | $N_{bunch}$ | $n_{bunch}$ | | | BS |
| Image current | | $N_{bunch}^2$ | $n_{bunch}$ | $\sigma_z^{-3/2}$ | $b^{-1}$ | BS |
| Photo-electron cloud | | $N_{bunch}^3$ | $n_{bunch}$ | | $b^{-2}$ | BS |
| Beam gas scattering | | $N_{bunch}$ | $n_{bunch}$ | | | CM |

Table 5: Parameters and specific beam-induced loads

| | LHC nom. | HE-LHC |
|---|---|---|
| Beam energy [TeV] | 7 | 16.5 |
| Bunch population [$10^{11}$ p] | 1.15 | 1.29 |
| Bunch number [-] | 2808 | 1404 |
| Bunch length [cm] | 7.55 | 6.55 |
| Beam-screen aperture radius [cm] | 2 | 1.3 |
| Synchrotron radiation [W/m] | 0.33 | 5.71 |
| Image current [W/m] | 0.36 | 0.44 |
| Photo-electron cloud [W/m] | 0.90 | 1.50 |
| Beam gas scattering [W/m] | 0.05 | 0.03 |

## OPERATING THE BEAM SCREENS AT A HIGHER TEMPERATURE

In addition to the nominal operating temperature of the beam screens at 46-20 K (range BS1), other possible temperature operating ranges compatible with the beam vacuum specification are 40-60 K (range BS2) or 85-100 K (range BS3). Increasing the operating temperature of the beam screen will have the following consequences:

- As the electrical resistivity of the copper on the beam-screen surface increases with the temperature, the image-current load will also increase proportionally. Measurements at 20 K, 50 K and 92.5 K on LHC beam-screen samples give a copper resistivity increase by factors 5.5 and 22 (see Figure 1). Consequently, the image current heat-load will increase from 0.44 to 2.4 and 9.8 W/m. A coating with HTS (like Bi-2223 or Y-123) may improve this figure dramatically. However, today this is a speculation.
- The temperature difference between the beam screen and the cold bore will increase, i.e. the heat inleaks on the cold mass will increase as well. Measurements on String 2 [2] indicate a heat inleak increase on the cold-mass of 0.17 and 0.71 W/m (see Figure 2).
- The present design of the LHC beam screen cooling loop based on a unit length of 53 m and two 3.7-mm inner-diameter capillaries per aperture is locally limited to a heat extraction of 2.4 W/m per aperture, i.e. 4.8 W per meter of machine. Changing the operating conditions and the specific heat load has a direct impact on the cooling capillary diameter. Table 6 gives the operating conditions of the beam screen cooling loops and the corresponding required capillary diameter assuming the same cooling loop configuration as today. The operation of the beam screen at 20 bar and between 40 and 60 K minimizes the cooling capillary diameter.

Table 6: Beam screen cooling capillary diameter

| | BS temperature range [K] | | |
| | BS1 | BS2 | BS3 |
|---|---|---|---|
| Inlet temperature [K] | 4.6 | 40 | 85 |
| Inlet pressure [bar] | 3.0 | 20 | 20 |
| Outlet temperature [K] | 20 | 60 | 100 |
| Outlet pressure [bar] | 1.3 | 18 | 18 |
| Specific heat load [W/m] | 7.65 | 9.45 | 16.3 |
| Loop length [m] | 50 | 50 | 50 |
| Nb of capillary per aperture | 2 | 2 | 2 |
| Capillary inner diam. [mm] | 4.4 | 3.8 | 6.0 |

## HEAT LOAD SUMMARY

Table 7 resumes the specific cryogenic heat load for the different temperature levels. Compared with the nominal LHC, depending on the beam-screen operating temperature range, the heat loads on the beam-screen circuits increase by a factor 4 to 9, and those on the cold-mass circuits by a factor 1.6 to 3.6.

Concerning the heat loads on the cold-mass circuits, the present LHC cooling loop is locally limited to 0.9 W/m, i.e. it is not compatible with the HE-LHC specific heat load corresponding to the 85-100 K operating temperature range (BS3) of the beam screens.

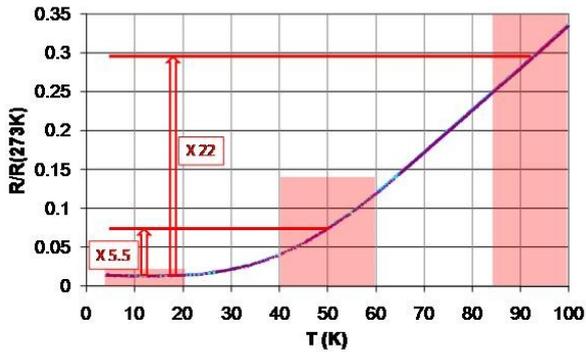

Figure 1: Copper resistivity

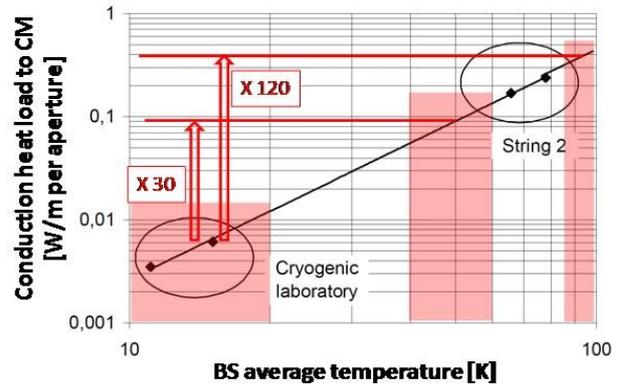

Figure 2: Conduction heat load to cold mass

Table 7: Cryogenic specific heat loads

| Temperature level | Heat load source | | LHC nominal | HE-LHC BS1 | BS2 | BS3 |
|---|---|---|---|---|---|---|
| TS | Heat inleaks | [W/m] | 7.7 | | 7.7 | |
| | **Total TS** | **[W/m]** | **7.7** | | **7.7** | |
| HI | Heat inleaks | [W/m] | 0.23 | | 0.23 | |
| | **Total HI** | **[W/m]** | **0.23** | | **0.23** | |
| BS | Heat inleaks | [W/m] | 0 | 0 | -0.17 | -0.71 |
| | Synchotron radiation | [W/m] | 0.33 | 5.71 | 5.71 | 5.71 |
| | Image current | [W/m] | 0.36 | 0.44 | 2.40 | 9.81 |
| | Photo-electron cloud | [W/m] | 0.90 | 1.50 | 1.50 | 1.50 |
| | **Total BS** | **[W/m]** | **1.82** | **7.65** | **9.45** | **16.3** |
| CM | Heat inleaks | [W/m] | 0.21 | 0.21 | 0.38 | 0.92 |
| | Resistive heating | [W/m] | 0.10 | 0.34 | 0.34 | 0.34 |
| | Beam-gas scattering | [W/m] | 0.05 | 0.03 | 0.03 | 0.03 |
| | **Total CM** | **[W/m]** | **0.36** | **0.58** | **0.74** | **1.29** |

## CONTINUOUS-CRYOSTAT COOLING CAPACITY

Assuming a continuous cryostat length of 2800 m and an overcapacity margin of 1.5, the required cooling capacity per continuous cryostat is given in Table 8 and is compared with the existing installed capacity of LHC sector cryogenic plants. Values in brackets correspond to the equivalent entropic capacity in kW at 4.5 K. Figure 3 shows the equivalent entropic capacity for the different temperature levels. Depending on the operating temperature range of the beam screen, the total equivalent entropic capacity of HE-LHC refrigerators varies from 31 to 19 kW at 4.5 K. Operating the beam screens between 4.6 and 20 K requires continuous-cryostat refrigerators about 1.7 times larger than the LHC sector refrigerators. Operating the beam screens between 40 and 60 K allows reducing the size of the continuous-cryostat refrigerators which becomes similar to the LHC sector refrigerators. Operating the beam screens between 85 and 100 K overloads the cold-mass temperature level. With a cold-mass operating temperature of 2 K, the optimum beam-screen temperature range is 40-60 K.

The electrical input power of the different scenarios, assuming a coefficient-of-performance of 250 W per W, is given in Table 9.

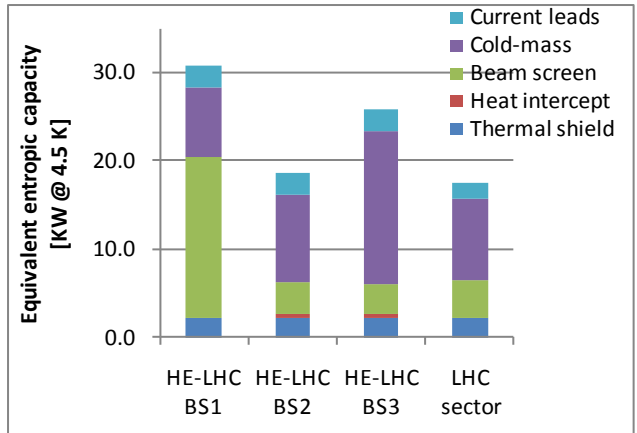

Figure 3: Equivalent entropic capacity

Table 8: Continuous cryostat cooling capacity per sector
(in brackets: equivalent entropic capacity in kW at 4.5 K)

| Temp. level | HE-LHC continuous cryostat | | | LHC high load sector |
|---|---|---|---|---|
| | BS1 | BS2 | BS3 | BS1 |
| TS [kW] | | 32 (2.2) | | 33 (2.2) |
| HI [kW] | 33 (18.4) | 1.0 (0.5) | 1.0 (0.5) | 7.7 (4.3) |
| BS [kW] | | 40 (3.5) | 69 (3.3) | |
| CM [kW] | 2.4 (7.8) | 3.1 (10) | 5.4 (17.4) | 2.7* (9.3) |
| CL [g/s] | | 56 (2.5) | | 41 (1.8) |
| Total | (30.8) | (18.6) | (25.8) | (17.6) |

*: 2.4 kW at 1.8 K plus 0.3 kW at 4.5 K

Table 9: Electrical input power for continuous-cryostat refrigerators

| Temp. level | HE-LHC CC refrigerator | | | LHC ref. |
|---|---|---|---|---|
| | BS1 | BS2 | BS3 | |
| Input power/refrigerator [MW] | 7.7 | 4.7 | 6.5 | 4.4 |
| Number of refrigerators [-] | 8 | 8 | 8 | 8 |
| Total input power [MW] | 62 | 37 | 52 | 35 |

## CONCLUSION

In these present cryogenic studies, no contingency has been introduced in the numbers. A lot of assumptions have to be confirmed like the splice resistance and number, the main magnet current, the current-leads distribution and number, and the cryostat performance.

The optimization of the refrigeration cycle has still to be done. Transient heat loads (ramp/de-ramp, fast de-ramp, quench), have still to be considered in order to define the correct level of buffering. In addition, the LSS loads have still to be considered with probably new cryoplants for insertions accommodating experiments.

Depending on the cooling scenario, up to 9 temperature levels have to be distributed along the continuous cryostats to supply or recover the different cooling loops. A rationalization study has to be done for reducing the number of distribution headers like operating the beam screen and the thermal shield with the same temperature range and/or cooling the resistive part of HTS current lead with a helium flow at a higher temperature and pressure (e.g. 40 K, 20 bar).

At this preliminary study phase, it is definitely too early to state on the possible reuse of LHC cryogenics. Nevertheless, it should be recalled that, at the end of the LHC (2030), the LHC cryogenics will be 30 to 40 years old. Taking into account the 20-year operation initially specified, major and wide overhauling has to be considered for the equipments which could be reused for the HE-LHC project (cryogenic plants, QRL, distribution boxes…).